\documentclass[onecolumn,showpacs,preprintnumbers,superscriptaddress,pre]{revtex4}
\usepackage{subfigure}
\usepackage{graphicx,dcolumn,bm,color,latexsym,amssymb}
\usepackage[latin1]{inputenc}

\definecolor{Blue}{rgb}{0.0,0.0,1}
\definecolor{Red}{rgb}{1,0.0,0.0}
\definecolor{Green}{rgb}{0,0.5,0.0}

\begin{document}

\title{Exact Nonequilibrium Work  Generating Function for a
Small Classical System}

\author{W. A. M. Morgado}
 \email{welles@fis.puc-rio.br}
\affiliation{Departamento de F\'{i}sica, Pontif\'{i}cia
Universidade Cat\'{o}lica\\ and National
Institute of Science and Technology for Complex
Systems\\ 22452-970, Rio de Janeiro, Brazil}

\author{D. O. Soares-Pinto} \email{diogo.osp@ursa.ifsc.usp.br}
\affiliation{Instituto de F\'{i}sica de S\~{a}o Carlos,
Universidade de S\~{a}o Paulo, P.O. Box 369, 13560-970, S\~{a}o
Carlos - SP, Brazil.}

\date{\today}

\begin{abstract}
We obtain the exact nonequilibrium work generating function
(NEWGF), for a small system consisting of a massive Brownian particle
connected to internal and external springs. The external work is provided to the system for  a finite time interval. The Jarzynski equality
(JE),  obtained in this case directly from the NEWGF, is shown  to be
valid for the present model, in an exact way regardless of the rate of external work.
\end{abstract}


\maketitle

%
%
\section{Introduction}
Due to the development of precision manipulation techniques at
very small length scales, such as atomic force microscopes or
optical tweezers, it has become possible to study the response of
small systems to applied external influences, such as the work
done by pulling apart the extremities of DNA
molecules~\cite{2004_PNAS_101_15038,2005_PhysToday_58_43,2005_Nature_437_231}. These nonequilibrium experiments can
yield equilibrium information (free energies) about the
system~\cite{2002_Science_296_1832,2005_Nature_437_231,
2009_PhysBiol_6_025011}. This can be obtained from relations such
as the Jarzynski equality~\cite{1997_PRL_78_2690,2000_PRE_61_2361} (JE).
The explicit form for the JE~\cite{1997_PRL_78_2690} relating the external work
fluctuations and the variation $\Delta F$ of the free energy
between these two states is
\begin{equation}
\overline{\exp\left\{-\beta\,W\right\}} =
\exp\left\{-\beta\,\Delta F\right\}. \label{Jarz}
\end{equation}

In order to understand the equality above, we observe that an important step in the derivation of Fluctuation Theorems, and the JE, is to make sure that the changes in the values of the Hamiltonian of the system are directly associated with the external work performed by the environment onto the system~\cite{2007_CRP_8_495,2007_JSTAT_P11002}. In fact, the distinction between environment and system is a matter of  choice. We can decide  which variables  are accounted for as belonging to the system, the remaining being part of the environment. In fact, conservative external forces might be incorporated into the Hamiltonian of the system as potential-energy terms~\cite{2007_CRP_8_495}, thus making the total Hamiltonian invariant under its effect.   So, for an isolated system the external work verifies ($x$ and $p$ represent all phase space variables)
\begin{equation}
W_{\mbox{external}}=\mathcal{H}(x_2,p_2,t_2)-\mathcal{H}(x_1,p_1,t_1),
\end{equation}
where positions and momenta evolve from state 1 to state 2 according to the dynamics of the system.
However, external forces arising from rheonomic constraints~\cite{livro_lanczos} might perform work on the system but, since they do not come from a simple expression of a conservative potential energy, they cannot be simply incorporated into the Hamiltonian as above. On the other hand, their work expression can be derived from the constraint's equations~\cite{livro_lanczos}. In fact, these  forces can do work on an otherwise isolated system and change the  Hamiltonian energy landscape upon which the phase-space point evolve in time.
\footnote{An example of such a potential is that of an instantaneously switched-on constant force$f(t)$~\cite{2008_PRL_100_020601}:
$$
\phi(x,t)=-f(t)\,x\equiv -f_0\,\Theta(t)\,x.
$$
For $t<0$, there is no potential energy, while for $t>0$ a potential energy $-f_0\,x$ arises~\cite{2007_CRP_8_495}. For instance, this can be thought as the energy of a very small bare electric charge placed in the inside of charged capacitor plates, with plates orthogonal to the x-axis. The charging of the capacitors happens almost instantly at $t=0$, and we assume that the plates were grounded at $t<0$.
In fact, the external work done by the force, described by the equation above, corresponds to the work done by the batteries in order to initially charge the capacitors up to a fixed voltage (part of that work shifts the particles potential energy and the remaining is used to create the electric field between the plates), producing a constant force upon the particle. It has the effect of shifting the initial potential energy of the charged particle by an amount $-f_0\,x_0$, where $x_0$ is the initial position of the charged particle (if we include the batteries within the system, then the potential energy they provide is taken to be internal). That external work is then given by
$$
W_{\mbox{ext},0} =  -f_0\,x_0 =  -\int_{t_1<0}^{t_2>0}dt\,f_0\,\dot\Theta(t)x_t,
$$
consistent with references~\cite{2007_CRP_8_495,2007_JSTAT_P11002}. The important point to be consistent with is that the external work is the one changing the energy landscape for the phase-space point. In contrast, if we keep the batteries as an external agent for all $t$, then the total external work is given by
$$
W_{\mbox{ext}} =  -f_0(x_{t_2}-x_{0})+W_{\mbox{ext},0}  =  -f_0\,x_{t_2}.
$$
The definition of what is the internal Hamiltonian, and what is external, is crucial as to which Work Fluctuation relation can be derived.~\cite{2007_CRP_8_495,2007_JSTAT_P11002}}
Thus, one has to be quite careful when defining the external work, as that choice can
substantially change the final form for any fluctuation
relation~\cite{2007_JSTAT_P11002}. Indeed, fluctuation relations that are derived under  distinct choices of the definition of work used will lead to distinct forms of the fluctuation theorems~\cite{2007_CRP_8_495,2007_JSTAT_P11002,1977_ZETF_72_238,1977_JETP_45_125, 1981_PHYSICA_106_443,1981_PHYSICA_106_480,2008_PRL_100_020601,2008_PRL_101_098901,
2008_PRL_101_098902,2008_PRL_101_098903,2008_PRL_101_098904,2007_PRE_75_011133} (FT).

The JE have been demonstrated exactly for systems initially thermalized at temperature $T$, that can be either mechanically closed or in contact with a thermostat (at temperature $T$ also) all the time. These results are obtained for systems that are large enough so that they may be placed into a true equilibrium state for which a valid partition function exists, and a correspondingly  valid free-energy associated with it~\cite{livro_huang}, according to the usual derivation of the canonical ensemble distribution. On the other hand, given the Hamiltonian for a small system, one can define its canonical partition function and calculate the corresponding free energy.

Several models~\cite{1999_cond-mat_9912121, 2005_EPL_69_643, 2006_JSMTE_3_5, 2007_PRL_99_068101, 2007_PRL_99_168101, 2007_JChemPhys_127_145105, 2008_JChemPhys_129_024102, 2009_JChemPhys_130_234116, 2010_PRL_104_090602} have been proposed where the JE is verified for distinct systems. In particular, for Ref.~\cite{1999_cond-mat_9912121}, the system consists of a particle pulled through a thermal bath in a manner that becomes time invariant as $t\rightarrow\infty$.  We propose the present model as a realization of a non-equilibrium process for a system that is clearly non-homogeneous in time. Irrespective of the duration of the process, the amount of external work done is finite, in average, and fluctuates around a well defined value.

Our model consists of a Hamiltonian that incorporates the kinetic energy term of a Brownian particle, and the potential energy terms associated with two springs connected to the particle: one represents a harmonic potential centered at $x=0$ an the other spring has one end fixed on the particle  while the other end is pulled, according to a given time protocol, as work is done on the system (mass and two springs) by the external constraint force at the pulling end of the  spring. This external force is  the only force that can change the energy of the system.  Our choice of definition for the work follows from the discussion above and seems well suited for studying the JE. This model can be thought as a prototype of a controllable heat engine that can operate in reversible or irreversible modes.

We calculated exactly the model dynamics, in the spirit of other models previously used by the authors~\cite{2009_PRE_79_051116,2008_PRE_77_011103}, and we analyzed the behavior of external work fluctuations  that are in contact with thermal
bath. The thermal bath is represented by a noise term in a
Langevin equation (LE)~\cite{livro_vankampen}. We approach that
problem from the point of view of a non-equilibrium work
generating function (NEWGF), equivalent to the complete
non-equilibrium work probability distribution.
Such functions have been used in the context of the JE
\cite{2005_PRE_72_046114, 2005_EuroLett_70_740}. They allow us to obtain the work
probability distribution for equilibrium and nonequilibrium
conditions.

Starting from the LE for an underdamped Brownian particle, we  obtain exact information on the
structure of the NEWGF. Solving exactly the LE for a system where the noise stochastic properties are known is
akin to solving the Kramers-Moyal equation~\cite{livro_vankampen}
for the probability distribution.  Establishing exact finite-time Langevin dynamics results gives us the possibility to calibrate analytical or numerical models via the JE,   in other words, it is a first principles calculation that corroborates the validity of JE, and it can also serve as a detailed testing ground for numerical simulation models.
The model consists of a Brownian particle of mass $m$, under the action of a harmonic potential $k$, and in contact with
a thermal reservoir at temperature $T$ and friction coefficient
$\gamma$. We attach to that particle an external spring
($k^{\prime}$), by one extremity, and pull the other extremity at a
fixed time rate (defining the work protocol). The
particle-reservoir coupling is represented by a Langevin force (noise)
$\eta(t)$. The external spring has the externally moving
extremity at the point $x_{spring}(t)=L(t)$, as work is externally
done into the system ($m$, $k$, $k^{\prime}$) without ambiguity:
the work is the product of the externally varying force applied to
the moving extremity of the spring $k^{\prime}$ with its
displacement $dL$. The model and protocol we use, varying an
external coordinate according to a pre-established time-rate, are
equivalent to others found in the
literature~\cite{2007_JSTAT_P11002}.

This paper is organized as follows. In Sec.II we define the model. In Sec.III we obtain the generating function for the work probability. In Sec.IV we derive the JE, followed by our conclusions in Sec.V.

%
%
\section{Model}
So, let us define our model LE:
\begin{eqnarray}
m\,\dot{v} & = & -\gamma\,v-k\,x
-k^{\prime}\,\left(x-L\right)+ {\eta}\label{1},\\
\dot{x} &=& v,\label{2}\\
L &=& L_0\,\left(1-e^{-t/\lambda}\right)\label{3}.
\end{eqnarray}

The process starts ($t=0$) with the system initially at
equilibrium with a reservoir (at temperature $T$). The initial
conditions ($x_0,v_0$) are distributed obeying the canonical
distribution at temperature $T$, and with $x_{spring}(t=0)=0$. Then, the external spring is moved [given $x_{spring}(t)=L(t)$] up
to $t=\tau$, which may or may not be extended to infinity. The
specific form of $L(t)$ given in Eq.(\ref{3}) was chosen for being
easy to manipulate, but it can be readily generalized. The rate
$\lambda$ can be set to any value, with $\lambda\rightarrow\infty$
corresponding to the reversible work process.

By taking the Laplace transform of
the Gaussian noise function's, we obtain for the second cumulant (given that the average of $\eta$ is null)
\begin{eqnarray}
\langle\tilde{\eta}(z_1)\tilde{\eta}(z_2)\rangle
&=&\frac{2\gamma T}{z_1+z_2}.
\end{eqnarray}
We can integrate this
system exactly by  techniques similar to the ones used previously~\cite{2009_PRE_79_051116, 2008_PRE_77_011103,
2006_PhysA_365_289}, where the integration paths are all described
therein. However, at present, we will use a direct solution technique which is different, and simpler than that in references~\cite{2009_PRE_79_051116, 2008_PRE_77_011103,
2006_PhysA_365_289}.

%
%
\section{Generating function for the external work}
In the spirit of the JE we define the precursor function to the NEWGF as
\begin{equation}
F(u)\equiv{\exp\left\{-i\,u\, W_{\tau}\right\}}=
\sum_{n=0}^{\infty}\frac{(-i\,u)^n}{n!}
{W_{\tau}^n},\label{NEWGF}
\end{equation}
where the average is taken over the thermal noise, which
corresponds to all possible nonequilibrium paths for the Brownian
particle. In order to construct the NEWGF we need to take averages, thermal (represented by $\left<f\right>$) and over the initial conditions (represented by $\overline{f}$). The expression for it reads
\begin{eqnarray*}
{\mathcal F}(u) & = & \left<\overline{F(u)}\right>\\ & = &  \sum_{n=0}^{\infty}\frac{(-iu)^n}{n!}\overline{\left<\left(\Delta\,U+\Delta\,W_p+\Delta\,W_h+\Delta\,W_{\eta}\right)^n\right>}\\
& = & \exp\left\{-iu\,(\Delta\,U+\Delta\,W_p)\right\}\, \sum_{n=0}^{\infty} \frac{(-iu)^{n}}{n!} \overline{\left(\Delta\,W_h\right)^{n}}\, \sum_{n=0}^{\infty} \frac{(-iu)^{n}}{n!}  \left<{\left(\Delta\,W_{\eta}\right)^{n}}\right>,
\end{eqnarray*}
where the partial terms for the external work $\Delta\,W_p$, $\Delta\,W_h$, and $\Delta\,W_{\eta}$ will be defined in the following.

The accumulated work-function, that measures the total external
work done on the system up to time $\tau$, $W_{\tau}$, is ($F_{\mbox{ext}}=-k^{\prime}\left(x(t)-L(t)\right)$)
\begin{eqnarray}
W_{\tau} &=&\int_{0}^{\tau}F_{ext}\,dL=-k^{\prime}\int_{0}^{\tau}dt \,
\frac{dL}{dt}\left(x(t)-L(t)\right)\nonumber\\
&=& \Delta\,U-k^{\prime}\int_{0}^{\tau}dt \,
\frac{dL(t)}{dt}x(t),\label{work}
\end{eqnarray}
with $\Delta\,U=k^{\prime}L^2(\tau)/2$. It is the
coupling of $x(t)$ and $\frac{\partial\,L(t)}{\partial\,t}$ will
give rise to the irreversible work loss.

A few thermodynamic properties for our system can be obtained
directly from the equilibrium partition function ${\it Z} =
\int_{-\infty}^{\infty} \frac{dp\,dx}
{\hbar}\,e^{-\beta\,\mathcal{H}(x,p)},$ where
$\beta\,\mathcal{H}(x,p)=\frac{1}{2T}\left(\frac{p^2}{m}+k
\,x^2+k^{\prime}\,(x-L)^2\right)$. We find $ F =
\left(\frac{kk^{\prime}}{k+k^{\prime}} \right)
\frac{L^2}{2}-T\ln\left(\frac{2\,\pi\,T}{\hbar}
\sqrt{\frac{m}{k+k^{\prime}}}\right),$  $S =  \ln
\left(\frac{2\,\pi\,T}{\hbar}
\sqrt{\frac{m}{k+k^{\prime}}}\right)+1,$ and $E = T +
\left(\frac{kk^{\prime}}{k+k^{\prime}} \right)\frac{L^2}{2}$,
where $T$ corresponds to the kinetic and  elastic energy
contributions around equilibrium (via equipartition theorem
$2\times\,T/2$). The second term on the RHS is the rest-energy of
two  springs, $k$ and $k^{\prime}$, of zero length, connected
serially with total extension $L$.

Keeping $T$ constant, the reversible work $W_r$ and the free-energy
$\Delta\,F$ are identical ($L(t=0)=0$):
\begin{equation}
W_r=\Delta F=  \left(\frac{kk^{\prime}}{k+k^{\prime}}
\right)\left(\frac{L^2}{2}\right)=
\left(\frac{k}{k+k^{\prime}}\right)\Delta\,U.\label{wrev}
\end{equation}
The dissipative work $W_d=W_{\tau}-W_r$ can be expressed as the
integral of fluctuations of $x$ around the instantaneous
equilibrium position
$x^{eq}=\frac{k^{\prime}}{k+k^{\prime}}\,L(t)$:
\begin{eqnarray}
W_d &=&-k^{\prime}\int_{0}^{\tau}dt \,
\frac{dL}{dt}\left(x(t)-\frac{k^{\prime}}{k+k^{\prime}}\,L(t)\right).\label{wdiss}
\end{eqnarray}

We can obtain a particular solution $x_S(t)$ associated with the source term $S(t)$, via the Green's function method, explicitly as
\begin{equation}
x_S(t) = \int_{0}^{t}dt'\,{\frac {2\,{{\rm e}^{-\frac{\theta\,(t-t')}{2}}}}{m\,\sqrt {4\,{\omega}^{2}-{\theta}^{2}}}}\, \sin\left(\frac{\sqrt {4
\,{\omega}^{2}-{\theta}^{2}}(t-t')}{2}\right)\,S(t'),
\end{equation}
where $\theta=\gamma/m$, and $\omega^2=(k+k^{\prime})/m$.

The exact solution $x(t)$ for Eqs.~\ref{1}-\ref{3}, taken in the sense that the exact form for the Langevin term $\eta(t)$ is known   (due to the linearity of the problem under scrutiny) is simply the sum of the homogeneous solution $x_h(t)$, the term $x_p(t)$ with source $L_0\,\left(1-e^{-t/\lambda}\right)$, and the term   $x_{\eta}(t)$ with source $\eta(t)$: $x(t)=x_h(t)+x_p(t)+x_{\eta}(t)$.

We may write the total work as
\begin{eqnarray}
W_{\tau} &=& \Delta\,U-k^{\prime}\int_{0}^{\tau}dt \,
\frac{\partial\,L(t)}{\partial\,t}(x_h(t)+x_p(t)+x_{\eta}(t))\nonumber\\&\equiv& \Delta\,U+ \Delta\,W_h^{\tau} +\Delta\,W_p^{\tau}+\Delta\,W_{\eta}^{\tau}.\label{WW}
\end{eqnarray}

The work expressions, for $\Delta\,W_{h}^{\infty}$ and $\Delta\,W_p^{\infty}$ are given below. The detailed form of the coefficients is given in Appendix~\ref{appb} in Eqs.~\ref{C1} and~\ref{C2}:
\begin{eqnarray}
\Delta\,W_{h}^{\tau} &=& C_1^{\tau}\,x_0+C_2^{\tau}\,v_0,\\
\Delta\,W_{\eta}^{\tau} &=&-k^{\prime}\int_{0}^{\tau}dt \,
\frac{\partial\,L(t)}{\partial\,t}x_{\eta}(t).
\end{eqnarray}

Making the summation for the NEWGF:
\begin{eqnarray*}
\sum_{n=0}^{\infty} \frac{(-iu)^{n}}{n!} \overline{\left(\Delta\,W_h\right)^{n}}
& = & \exp\left\{-u^2\frac{C_1^{\tau\,2}\,\overline{x_0^2}+C_2^{\tau\,2}\,\overline{v_0^2}}{2}\right\},
\end{eqnarray*}
where $\overline{x_0^2}=\frac{T}{m\omega^2}$ and $\overline{v_0^2}=\frac{T}{m}$.

The thermal contribution for the work arises from the need to compensate the dissipative coupling of the noise term with the pulling rate expressed below:
\begin{eqnarray*}
\Delta\,W_{\eta}^{\tau} &=&-k^{\prime}\int_{0}^{\tau}dt \,
\frac{\partial\,L(t)}{\partial\,t}x_{\eta}(t).
\end{eqnarray*}
The contribution above is the only source of irreversibility into the system. The expression for $\Delta\,W_{\eta}^{\tau}$ is then:

\begin{eqnarray}
\Delta\,W_{\eta}^{\tau}
&=&-\lim_{\epsilon\rightarrow0}\int_{-\infty}^{\infty}\frac{dq_1}{2\pi} \,\tilde{\eta}(iq_1+\epsilon)\,
\times\nonumber\\
&\times&\left(\frac{2\,L_0\,k^{\prime}}{\lambda \,m}\right)\int_{0}^{\tau}dt \int_{0}^{t}dt' \,{\frac {{{\rm e}^{(iq_1+\epsilon)t'-\frac{t}{\lambda}-\frac{\theta\,(t-t')}{2}}} }{\sqrt {4\,{\omega}^{2}-{\theta}^{2}}}} \, \sin\left(\frac{\sqrt {4
\,{\omega}^{2}-{\theta}^{2}}(t-t')}{2}\right) \nonumber\\
& = & -\lim_{\epsilon\rightarrow0}\,\int_{-\infty}^{\infty}\frac{dq}{2\pi}\,\tilde{\eta}(iq+\epsilon)\,\left(W_{\eta1}^{\tau}(iq+\epsilon)+ W_{\eta2}^{\tau}(iq+\epsilon)+W_{\eta3}^{\tau}(iq+\epsilon)+W_{\eta4}^{\tau}(iq+\epsilon)\right),
\end{eqnarray}
where the detailed expressions for the $W_{\eta j}^{\tau}$ listed in Appendix~\ref{appb}.

The Gaussian property of the noise function leads to the disappearing of the odd momenta of $\Delta\,W_{\eta}^{\tau} $. The even ones are given as products of $\left<\Delta\,W_{\eta}^{\tau} \,\,\Delta\,W_{\eta}^{\tau} \right>$, that can be expressed as  sums of terms of the form:
\begin{eqnarray}
I_{ij} & = & \lim_{\epsilon\rightarrow0}\,\int_{-\infty}^{\infty}\frac{dq_1}{2\pi}\,\int_{-\infty}^{\infty}\frac{dq_2}{2\pi}\, \left<\tilde{\eta}(iq_1+\epsilon)\,\tilde{\eta}(iq_2+\epsilon)\right>\,W_{\eta i}^{\tau}(iq_1+\epsilon)\, W_{\eta j}^{\tau}(iq_2+\epsilon)\nonumber\\
& = & \lim_{\epsilon\rightarrow0}\,\int_{-\infty}^{\infty}\frac{dq_1}{2\pi}\,\int_{-\infty}^{\infty} \frac{dq_2}{2\pi}\,\frac{2\,\gamma\,T}{(iq_1+ iq_2+2\epsilon)}\,W_{\eta i}^{\tau}(iq_1+\epsilon)\, W_{\eta j}^{\tau}(iq_2+\epsilon).
\end{eqnarray}
There are ten possible distinct pairs above but we can show that only $I_{11}$, $I_{12}$,  and $I_{11}$ give non-zero results. Thus,  we have
\begin{eqnarray*}
\left<(\Delta\,W_{\eta}^{\tau})^{2n} \right> & = &  \frac{(2n)!}{n!\,2^n}\left(I_{11}+2I_{12}+I_{22}\right)^n,
\end{eqnarray*}
yielding
\begin{eqnarray*}
\sum_{n=0}^{\infty} \frac{(-iu)^{n}}{n!}  \left<{\left(\Delta\,W_{\eta}\right)^{n}}\right>&=&\exp\left\{-u^2\,\frac{I_{11}+2I_{12}+I_{22}}{2}\right\}.
\end{eqnarray*}

The expression for the NEWGF  reads
\begin{eqnarray*}
{\mathcal F}(u) & = &  \exp\left\{-iu\,(\Delta\,U+\Delta\,W_p) -u^2\frac{C_1^{\tau\,2}\,\overline{x_0^2}+C_2^{\tau\,2}\,\overline{v_0^2}}{2}-u^2\,\frac{I_{11}+2I_{12}+I_{22}}{2} \right\}\,\equiv \, \exp\left\{-iu\,{\mathcal R}_1 -u^2{\mathcal R}_2 \right\}.
\end{eqnarray*}
Generating functions of Gaussian shape~\cite{2005_EPB_43_521,2007_PRE_76_050101,2007_JSP_126_1}, for quadratic time-varying potentials, have already been found in the literature. The explicit expressions for ${\mathcal R}_1$ and ${\mathcal R}_2$ can be found in Appendix~\ref{appb}.

%
%
\section{The Jarzynski Equality}
The Jarzynski equality (JE) is verified, after some tedious, but straightforward, manipulations of the terms ${\mathcal R}_1$ and ${\mathcal R}_2$ from Appendix~\ref{appb}, for all values of  $\tau$ and $\lambda$ since for $u=-i/T$,
\begin{equation}
{\mathcal F}\left(-\frac{i}{T}\right) = \left<\exp\left(-\frac{W}{T}\right)\right>
 =  \exp\left(-\frac{{\mathcal R}_1}{T}+\frac{{\mathcal R}_2}{T^2} \right)= \exp\left(-\frac{\Delta\,F}{T} \right),
\end{equation}
where $\Delta\,F= \frac{kk^{\prime}}{(k+k^{\prime})}L_{f}^2$, and $L_{f}=L(\tau)$.
Despite the highly non trivial dependence of  ${\mathcal R}_1$, and ${\mathcal R}_2$, on $\lambda$ and $\tau$, at $u=-i/T$ the correct cancellations occur and the JE is verified.

We notice that by fixing the final position of the external spring $0<L_f\leq L_0$ the ratio $\tau/\lambda$ gets fixed,
\[
\frac{\tau}{\lambda}= \ln\left(\frac{L_0}{L_0-L_f}\right),
\]
which gives us an infinite number of distinct protocols for taking the system from state $A\equiv{(L=0)}$ to state $B\equiv{(L=L_f)}$, verifying the JE for all cases for a fixed $\frac{\Delta\,F}{T} = \frac{F_B-F_A}{T} $.

The Gaussian form of ${\mathcal F}(u)$ is the expected one due to the linear form of the harmonic potential~\cite{1982_PhysRep_88_207}. The present model is an explicit dynamic solution that could be extended to other forms of the noise, in the case that its cumulants are known.  In fact, for the quasi-static case,  $\lambda,\tau\rightarrow\infty$ ($\tau/\lambda$ fixed), ${\mathcal F}(u) \rightarrow \exp\left\{-i\,u\,\Delta F\right\}\rightarrow p(W)=\delta(W-\Delta F)$. In this case, there is only one way of carrying out the process.

We can also obtain the forward ratio for the work distributions.
We notice that $W$ is finite (with probability equals to 1) and
positive. The probability distribution for the total work $p(W)$
is the inverse Fourier transform of $F_W(u)$:
\begin{equation}
p(W) = \int_{-\infty}^{\infty}\frac{du}{2\pi}\,{\cal F}(u)\,e^{iuW}.
\end{equation}

The expression for $p(W)$ explicitly reads:
\begin{eqnarray}
p(W)
&=&
\sqrt{\frac{\pi}{\mathcal{R}_2}}\exp\left\{-\frac{(W-\mathcal{R}_1)^2}{4\,\mathcal{R}_2}\right\}.
\end{eqnarray}
It can be seen that the average work done externally on the system is given ${\mathcal R}_1$, while the variance is given by $2{\mathcal R}_2$, which is proportional to $T$ for all $\lambda$ and $\tau$.

The ratio $p(W)/p(-W)$, not to be mistaken with the Crooks Fluctuation expression~\cite{2000_PRE_61_2361},  can be obtained explicitly
\begin{eqnarray}
\frac{p(W)}{p(-W)}
&=& \exp\left\{\frac{W \, \mathcal{R}_1}{4\, \mathcal{R}_2^2}\right\}.
\end{eqnarray}

The expression above is well behaved in the instantaneous work case, $\lambda\rightarrow0$, since averaging over an ensemble of initial conditions distributed with temperature $T$ converges, while it becomes proportional to a delta-function when $\lambda\rightarrow\infty$, since ${\mathcal{R}_2}\rightarrow0$ in this case. In fact, the averages of the work do depend on the work rate $\lambda$ and the elapsed time $\tau$, but the JE arises regardless of it.

%
%
\section{Conclusions}
In conclusion, we develop an exact technique appropriate for
treating a system consisting of a massive particle coupled to two harmonic
springs in contact with an external thermal reservoir, represented
by a Langevin force term. External work can be done by pulling one
of the springs at a given rate, which is the protocol we follow.
The main advantage of this model is that we can explicitly make
all the calculations with no approximations. No approximations are needed in respect to the mass of the particle, the calculations being able to take care of the particle's inertia exactly. This model can be thought as a controllable, and simple, heat engine.

To the best of our knowledge, for the first a Langevin model was exactly integrated, taking into account inertia and general initial conditions, for a Brownian particle under the action of a given protocol.  An exact form for the nonequilibrium work generating function (NEWGF) is obtained. The Jarzynski equality is then explicitly verified, such as predicted~\cite{1997_PRL_78_2690},  showing that the method used in this paper can be seen as a first principle exact, and nontrivial, verification of the JE. This shows the appropriateness of using white Gaussian noise to represent the interaction between a thermal bath and a  system.
Furthermore, the work probability distribution is derived explicitly for this case and shows that the moments of the work $W$ are complex functions of the work rate $\lambda$ and the time interval $\tau$.

\acknowledgements

The authors would like to thank C. Jarzynski for useful discussions.
W.A.M.M. would like to thank the Brazilian funding agencies Faperj, CAPES
and CNPq, and D.O.S.P. would like to thank the Brazilian funding
agency FAPESP for the financial support.


\appendix %
\setcounter{equation}{0}
%
%
%
%
\section{Finite-time coefficients}\label{appb}
The exact finite time coefficients  are listed below.
\begin{eqnarray}
C_{1}^{\tau} &=& - \frac{ k^{\prime} L_{{0}} \left( 2\,\lambda\,{\omega}^{2}-{\theta}^{2}
\lambda-\theta \right) }{\sqrt {4\,{\omega}^{2}-{\theta}^{2}}
\left( 1+\theta\,\lambda+{\lambda}^{2}{\omega}^{2} \right) } \,{{\rm e}^{-{\frac {\tau\, \left( 2+\theta\,\lambda
 \right) }{2\,\lambda}}}} \sin \left( \frac{\tau\,\sqrt {4\,{\omega}^{2}-{\theta}^{2}}}{2} \right)\nonumber \\
&+& \frac{ k^{\prime} L_{{0}} \left( 1+\theta\,\lambda \right)}{  \left( 1+
\theta\,\lambda+{\lambda}^{2}{\omega}^{2} \right) } \,{{\rm e}^{-{\frac {\tau\, \left( 2+\theta\,\lambda
 \right) }{2\,\lambda}}}} \cos \left( \frac{\tau\,\sqrt {4\,{\omega}^{2}-{\theta}^{2}
}}{2} \right) \nonumber\\
&-&{\frac {k^{\prime} L_{{0}} \left( 1+\theta\,\lambda \right) }{1+\theta\,\lambda+{\lambda
}^{2}{\omega}^{2}}} \label{C1}
\end{eqnarray}
\begin{eqnarray}
C_{2}^{\tau} &=& {\frac { k^{\prime}L_{{0}} \left( 2+\theta\,\lambda \right)
}{\left( 1+\theta\,\lambda+{
\lambda}^{2}{\omega}^{2} \right) \sqrt {4\,{\omega}^{2}-{\theta}^{2}}}} \,{{\rm e}^{-{\frac {\tau\, \left( 2+\theta\,\lambda \right) }{2\,\lambda}}}} \sin \left( \frac{\tau\,\sqrt {4\,{\omega}^{2}-{\theta}^{2}}}{2} \right)\nonumber \\
&+& \frac{ k^{\prime}L_{{0}} \lambda}{  \left( 1+\theta\,\lambda+{\lambda}^{2}{\omega}^{2} \right) } \,{{\rm e}^{-{\frac {\tau\, \left( 2+\theta\,\lambda
 \right) }{2\,\lambda}}}} \cos \left( \frac{\tau\,\sqrt {4\,{\omega}^{2}-{\theta}^{2}
}}{2} \right) \nonumber \\
&-& {\frac {k^{\prime}\lambda\,L_{{0}}}{1+\theta\,\lambda+{\lambda}^{2}{\omega
}^{2}}}\label{C2}
\end{eqnarray}
\begin{eqnarray}
\Delta W_{p}^{\tau} &=&- \frac {\theta\,{k^{\prime}}^{2}{L_{{0}}}^{2}\,\left( 3\,{\omega}^{2}{\lambda}^{2}-{\theta}^{2}{\lambda}^{2}+1
 \right)}
{m\,{\omega}^{2}\,\sqrt {4\,{\omega}^{2}-{\theta}^{2}} \left[\left(1 + \lambda^2 \omega^2 \right)^2 -\theta^2 \lambda^2\right]}\,\,{{\rm e}^{-{\frac {\tau\, \left( 2+\theta\,\lambda \right) }{2\,\lambda}}}} \sin \left( \frac{\tau\,\sqrt {4\,{\omega}^{2}-{\theta}^{2}}}{2} \right) \nonumber \\
&-& \frac{{k^{\prime}}^{2}{L_{{0}}}^{2} \left( -{\theta}^{2}{\lambda}^{2}+{\omega}^{2}{\lambda}^{2}+1
 \right)}{m\,{\omega}^{2}
 \left[\left(1 + \lambda^2 \omega^2 \right)^2 -\theta^2 \lambda^2\right]} \,{{\rm e}^{-{\frac {\tau\, \left( 2+\theta\,\lambda
 \right) }{2\,\lambda}}}} \cos \left( \frac{\tau\,\sqrt {4\,{\omega}^{2}-{\theta}^{2}
}}{2} \right) \nonumber \\
&-& \frac{{k^{\prime}}^{2}{L_{{0}}}^{2} \left( -2\,{\omega}^{2}{\lambda}^{2}+{\omega}^{2}{\lambda}^{2}{
{\rm e}^{{\frac {\tau}{\lambda}}}}-2\,\theta\,\lambda-2 \right)}{2\,m\,{\omega}^{2} \left( 1+\theta\,\lambda+{\omega}^{2}{\lambda}^{2} \right)} \,{{\rm e}^{-{\frac {\tau}{\lambda}}}} \nonumber \\
&-&\frac{{k^{\prime}}^{2
}{L_{{0}}}^{2}{\lambda}^{2}}{2\,m\left( 1-\theta\,\lambda+{\omega}^{2}{\lambda}^
{2} \right)}\,{{\rm e}^{-{\frac {2\,\tau}{\lambda}}}}
 \label{Wht}
\end{eqnarray}
\begin{equation}
W_{\eta1}^{\tau}(s) = \frac{k^{\prime}\,L_{{0}}}{m \left( \theta\,s+{s}^{2}+{\omega}^{2} \right)  \left( s\lambda-
1 \right) } \,{\rm e}^{{\frac {\tau\, \left( s\lambda-1 \right) }{\lambda}}}
\end{equation}
\begin{equation}
W_{\eta2}^{\tau}(s) =  -{\frac {k^{\prime}\,L_{{0}}{\lambda}^{2}}{ m\left( s\lambda-1 \right)  \left(
\theta\,\lambda+1+{\omega}^{2}{\lambda}^{2} \right) }}
\end{equation}
\begin{equation}
 W_{\eta3}^{\tau}(s) = - \frac {\left( -\theta\,\lambda\,s-{\theta}^{2}\lambda+2\,\lambda\,{
\omega}^{2}-2\,s-\theta \right)k^{\prime}\, L_{{0}}}{m\sqrt {4\,{\omega}^{
2}-{\theta}^{2}} \left( \theta\,s+{s}^{2}+{\omega}^{2} \right)
 \left( \theta\,\lambda+1+{\omega}^{2}{\lambda}^{2} \right)  } \,{{\rm e}^{-{\frac {\tau\, \left( 2+\theta\,\lambda \right) }{2\,\lambda}}}} \sin \left( \frac{\tau\,\sqrt {4\,{\omega}^{2}-{\theta}^{2}}}{2} \right)
\end{equation}
\begin{equation}
 W_{\eta4}^{\tau}(s) = \frac{ \left( s\lambda+\theta\,\lambda+1 \right) k^{\prime}\,L_{{0}}}{ m\left(
\theta\,s+{s}^{2}+{\omega}^{2} \right)  \left( \theta\,\lambda+1+
{\omega}^{2}{\lambda}^{2} \right)} \,{{\rm e}^{-{\frac {\tau\, \left( 2+\theta\,\lambda \right) }{2\,\lambda}}}} \cos \left( \frac{\tau\,\sqrt {4\,{\omega}^{2}-{\theta}^{2}}}{2} \right)
\end{equation}
\begin{eqnarray}
I_{11} &=& {\frac {\gamma\,T{L_{{0}}}^{
2}{k^{\prime}}^{2} \left( -\theta\,\lambda-1+\lambda\,\omega \right)  \left( 1+\theta\,
\lambda+\lambda\,\omega \right)}{ {m}^{2}{\omega}^{2}
\sqrt {4\,{\omega}^{2}-{\theta}^{2}} \left( 1+
\theta\,\lambda+{\omega}^{2}{\lambda}^{2} \right)^{2}}} \,{{\rm e}^{-{\frac {\tau\, \left( 2+\theta\,\lambda \right) }{\lambda}}}} \sin \left( \tau\,\sqrt {4\,{\omega}^{
2}-{\theta}^{2}} \right)  \nonumber \\
&-& \frac{\gamma\,T{L_{{0}}}^{2}{k^{\prime}}^{2} \left( -{\theta}^{3}{\lambda}^{2}-\theta-2\,{\theta}^{2}\lambda+4\,
\lambda\,{\omega}^{2}+3\,\theta\,{\lambda}^{2}{\omega}^{2} \right)}{{m}^{2}{\omega}^{2} \left( 4\,{\omega}^{
2}-{\theta}^{2} \right) \left( 1+\theta\,\lambda+{\omega}^{2}{
\lambda}^{2} \right)^{2}} \,{{\rm e}^{-{\frac {\tau\, \left( 2+\theta\,\lambda \right) }{\lambda}}}} \cos \left( \tau\,\sqrt {4\,{\omega}^{2}-{\theta}^{2}} \right)  \nonumber \\
&-& \frac{ \gamma\,T{L_{{0}}}^{2}{k^{\prime}}^{2} \left( {\theta}^{3}\lambda-4\,\theta\,\lambda\,{\omega}^{2}-4\,{
\omega}^{2}+4\,{{\rm e}^{-\tau\,\theta}}{\omega}^{2}+{\theta}^{2}
 \right) }{ {m}^{2}{\omega}^{2} {\theta} \left( 4\,{\omega}^{2}-{\theta}^{2} \right) \left( 1+\theta\,\lambda+{\omega}^{2}{\lambda}^{2}
 \right)} \,{{\rm e}^{-{\frac {2\,\tau
}{\lambda}}}}
\end{eqnarray}
\begin{eqnarray}
I_{12} &=&  \frac{2\,\gamma\,T{L_{{0}}}^{2}{k^{\prime}}^{2}{\lambda}^{2} \left( {\theta}^{2
}{\lambda}^{2}-2\,{\omega}^{2}{\lambda}^{2}-2 \right)}{{m}^{2}
{\sqrt {4\,{\omega}^{2}-{\theta}^{2}}} \left( 1+\theta\,
\lambda+{\omega}^{2}{\lambda}^{2} \right) \left[\left(1 + \lambda^2 \omega^2 \right)^2 -\theta^2 \lambda^2\right]} \,{{\rm e}^{-{\frac {\tau\, \left( 2+\theta\,\lambda \right) }{2\,\lambda}}}} \sin \left( \frac{\tau\,\sqrt {4\,{\omega}^{2}-{\theta}^{2}}}{2} \right)\nonumber \\
&+& \frac{2\,\gamma\,T{L_{{0}}}^{2}{k^{\prime}}^{2}{\lambda}^{4}\theta}{{m}^{2} \left[\left(1 + \lambda^2 \omega^2 \right)^2 -\theta^2 \lambda^2\right]} \,{{\rm e}^{-{\frac {\tau\, \left( 2+\theta\,\lambda \right) }{2\,\lambda}}}} \cos \left( \frac{\tau\,\sqrt {4\,{\omega}^{2}-{\theta}^{2}}}{2} \right)\nonumber \\
&-& \frac{\gamma\,T{L_{{0}}}^{2}{k^{\prime}}^{2}{\lambda}^{3}}{{m}^{2} \left[\left(1 + \lambda^2 \omega^2 \right)^2 -\theta^2 \lambda^2\right]} \,{{\rm e}^{-{\frac
{2\,\tau}{\lambda}}}}
\end{eqnarray}
\begin{equation}
I_{22} = {\frac {\gamma\,T{L_{{0}}}^{2}{k^{\prime}}^{2}{\lambda}^{3}}{ \left( 1+
\theta\,\lambda+{\omega}^{2}{\lambda}^{2} \right) ^{2}{m}^{2}}}
\end{equation}
\begin{eqnarray}
{\mathcal R}_1 &=& - \frac{{k^{\prime}}^{2}{L_{{0}}}^{2}\left( 1+{\omega}^{2}{\lambda}^{
2}-{\lambda}^{2}{\theta}^{2} \right)}{m {\omega}^{2} \left[\left(1+{\omega}^{2}{\lambda}^{2}\right)^{2} - {\theta}^{2}{\lambda}^{2} \right]}
\,{{\rm e}^{-{\frac {\tau\, \left( 2+\theta\,\lambda \right) }{2\,\lambda}}}} \cos \left( \frac{\tau\,\sqrt {4\,{\omega}^{2}-{\theta}^{2}}}{2} \right) \nonumber\\
&-& \frac{{k^{\prime}}^{2}{L_{{0}}}^{2}\theta \left( 1-{\lambda}^{2}
{\theta}^{2}+3\,{\omega}^{2}{\lambda}^{2} \right)}
{m {\omega}^{2} {\sqrt {4\,{\omega}^{2}-{\theta}^{2}}}  \left[\left(1+{\omega}^{2}{\lambda}^{2}\right)^{2} - {\theta}^{2}{\lambda}^{2} \right]}
\,{{\rm e}^{-{\frac {\tau\, \left( 2+\theta\,\lambda \right) }{2\,\lambda}}}} \sin \left( \frac{\tau\,\sqrt {4\,{\omega}^{2}-{\theta}^{2}}}{2} \right)\nonumber\\
&-& \frac{{L_{{0}}}^{2}k^{\prime} \omega^2 \left[k^{\prime} \lambda^2 \left(1 + \theta \lambda\right) +
m \left(-2 + \theta^2 \lambda^2 - 2 \lambda^2 \omega^2 - \lambda^4 \omega^4 \right) \right]}
{2 m {\omega}^{2} \left[\left(1+{\omega}^{2}{\lambda}^{2}\right)^{2} - {\theta}^{2}{\lambda}^{2} \right]} \,{{\rm e}^{-{\frac {2\,\tau}{\lambda}}}} \nonumber \\
&-& \frac{{L_{{0}}}^{2}k^{\prime} \left[-k^{\prime\,2}
+ k^{\prime} \lambda^2 \left(\theta^2 - \lambda^2 \omega^4\right)
+ \omega^2 \left( m - m \theta^2 \lambda^2
+ 2 m \lambda^2 \omega^2 + \lambda^4 \omega^4\right)\right]}
{m {\omega}^{2} \left[\left(1+{\omega}^{2}{\lambda}^{2}\right)^{2} - {\theta}^{2}{\lambda}^{2} \right]} \,{{\rm e}^{-{\frac {\tau}{\lambda}}}} \nonumber \\
&+& \frac{{L_{{0}}}^{2}k^{\prime} {\omega}^2 \left(1 - \theta \lambda + {\lambda}^2 \omega^2 \right) \left(m +
m \theta \lambda - k \lambda^2 + m \lambda^2 \omega^2\right)}
{2 m {\omega}^{2} \left[\left(1+{\omega}^{2}{\lambda}^{2}\right)^{2} - {\theta}^{2}{\lambda}^{2} \right]}
\end{eqnarray}
\begin{eqnarray}
{\mathcal R}_2 &=& - \frac{\left( 1+{\omega}^{2}{\lambda}^{2}-{\lambda}^{2}{\theta}^{2}
 \right) T{L_{{0}}}^{2}{k^{\prime}}^{2}}{m {\omega}^{2}
\left[\left(1+{\omega}^{2}{\lambda}^{2}\right)^{2} - {\theta}^{2}{\lambda}^{2} \right]} \,{{\rm e}^{-{\frac {\tau\, \left( 2+\theta\,\lambda \right) }{2\,\lambda}}}} \cos \left( \frac{\tau\,\sqrt {4\,{\omega}^{2}-{\theta}^{2}}}{2} \right)\nonumber\\
&-& \frac{ \theta  T{L_{{0}}}^{2}{k^{\prime}}^{2}\left( 1-{\lambda}^{2}{\theta}^{2}+3\,{\omega}^{2}{\lambda}^{2}
 \right)}{m {\omega}^{2} {\sqrt {4\,{\omega}^{2}-{\theta}^{2}}} \left[\left(1+{\omega}^{2}{\lambda}^{2}\right)^{2} - {\theta}^{2}{\lambda}^{2} \right]
}
\,{{\rm e}^{-{\frac {\tau\, \left( 2+\theta\,\lambda \right) }{2\,\lambda}}}} \sin \left( \frac{\tau\,\sqrt {4\,{\omega}^{2}-{\theta}^{2}}}{2} \right)\nonumber\\
&+& \frac{T{k^{\prime}}^{2}{L_{{0}}}^{2} \left( 1+{\omega}^{2}{\lambda}^{2}
-\theta{\omega}^{2}{\lambda}^{3}-{\theta}^{2}{\lambda}^{2} \right)}
{2 m {\omega}^{2} \left[\left(1+{\omega}^{2}{\lambda}^{2}\right)^{2} - {\theta}^{2}{\lambda}^{2} \right]}\,{{\rm e}^{-
{\frac {2\,\tau}{\lambda}}}} \nonumber \\
&+& \frac{T{k^{\prime}}^{2}{L_{{0}}}^{2} \left( 1+\theta\,{\lambda}^{3}{\omega
}^{2}+{\omega}^{2}{\lambda}^{2}-{\lambda}^{2}{\theta}^{2}\right)}
{2 m {\omega}^{2} \left[\left(1+{\omega}^{2}{\lambda}^{2}\right)^{2} - {\theta}^{2}{\lambda}^{2} \right]}
\end{eqnarray}

\end{document}